\documentclass[pdflatex,sn-mathphys-num]{sn-jnl}

\usepackage{graphicx}%
\usepackage{multirow}%
\usepackage{amsmath,amssymb,amsfonts}%
\usepackage{amsthm}%
\usepackage{mathrsfs}%
\usepackage[title]{appendix}%
\usepackage{xcolor}%
\usepackage{textcomp}%
\usepackage{manyfoot}%
\usepackage{booktabs}%
\usepackage{algorithm}%
\usepackage{algorithmicx}%
\usepackage{algpseudocode}%
\usepackage{listings}%
 \usepackage{ulem}
\usepackage{kotex}%
\usepackage{braket}

\theoremstyle{thmstyleone}%
\theoremstyle{thmstyletwo}%

\theoremstyle{thmstylethree}%

\begin{document}

\title[Article Title]{Passive Polarization Stabilization for Robust Entanglement Distribution via Cross-Aligned Polarization Maintaining Fiber Pairs}


\author[]{\fnm{Jin-Woo} \sur{Kim}}
\equalcont{These authors contributed equally to this work.}

\author[]{\fnm{Minchul} \sur{Kim}}
\equalcont{These authors contributed equally to this work.}

\author*[]{\fnm{Jiho} \sur{Park*}}\email{jiho5329@etri.re.kr}

\author[]{\fnm{Junsang} \sur{Oh}}

\author[]{\fnm{Kyongchun} \sur{Lim}}

\author[]{\fnm{Byung-Seok} \sur{Choi}}

\author[]{\fnm{Chun Ju} \sur{Youn}}

\affil[]{\orgname{Electronics and Telecommunications Research Institute (ETRI)}, \orgaddress{\city{Daejeon}, \postcode{34129}, \country{Republic of Korea}}}




\abstract{
Maintaining stable entanglement distribution through perturbed fiber links is essential for practical quantum-optics experiments, yet it remains challenging because of polarization fluctuations and phase or temporal-delay variations. We demonstrate stable entangled-photon transmission using a cross-aligned polarization-maintaining fiber (CAPMF) structure composed of two polarization-maintaining fiber sections with mutually orthogonal principal axes. The CAPMF configuration passively compensates polarization fluctuations without real-time active polarization control. We theoretically analyze the CAPMF structure and experimentally verify its stabilization performance under external mechanical perturbations. In the experiment, the single-mode fiber configuration yields an average visibility of $0.7655$ and a CHSH value of $S=1.7714$, whereas the CAPMF configuration maintains an average visibility of $0.9843$ and a CHSH value of $S=2.6838$. These results show that CAPMF offers a simple and robust architecture for stabilizing fiber-interface sections in practical entanglement-distribution systems.
}




\maketitle

\section{Introduction}\label{sec1}
Quantum entanglement is a central resource in quantum information technologies, enabling nonclassical correlations that have no analogue in classical systems. It plays an essential role in a broad range of applications, including quantum key distribution (QKD)~\cite{bennett1992quantum, ekert1991quantum}, quantum networks~\cite{kimble2008quantum}, and distributed quantum computing~\cite{cirac1999distributed}. In particular, entanglement distribution enables spatially separated nodes to share quantum states and forms the basis of quantum communication networks, supporting applications such as quantum direct communication~\cite{zhang2017quantum}, quantum repeaters~\cite{li2019experimental}, and large-scale quantum networks~\cite{chen2021integrated, lim2011experimental}. For these applications, the distributed entanglement must remain stable under realistic transmission conditions. 

Among the available physical platforms, photons are particularly attractive for entanglement distribution because of their weak interaction with the environment and their suitability for long-distance transmission~\cite{yin2017satellite}. Nevertheless, weak interaction does not imply complete immunity to environmental perturbations. In practical transmission systems, optical loss, polarization drift, and phase fluctuations can still degrade the quality of distributed entanglement. Accordingly, maintaining polarization and phase stability during entanglement transmission is essential for practical quantum communication and distributed quantum information processing~\cite{kim2024fully, Seong2026}.

Photonic entanglement distribution can be implemented through either fiber-based links or free-space links. Fiber transmission provides stable optical paths and high compatibility with existing optical communication infrastructure, but conventional single-mode fiber (SMF) does not intrinsically preserve polarization because environmental perturbations can cause the fiber birefringence. Free-space transmission has therefore been actively investigated for satellite-based entanglement distribution and long-distance quantum communication involving mobile nodes~\cite{yin2017satellite, ursin2007entanglement, tian2024experimental, nauerth2013air, Kim2024FreeSpaceQKD_WDM, liu2021optical, liu2020drone}. Nevertheless, practical free-space systems often still rely on short fiber interfaces to connect the entangled-photon source, transmitting optics, receiving optics, and measurement modules~\cite{yin2017satellite, yin2020entanglement, Kim2024HighBrightnessCHSH, liu2021optical, liu2020drone}. Because these fiber sections are much shorter than the main transmission channel, polarization and phase stability can be more critical than transmission loss. In mobile or field-operated platforms, bending, mechanical vibration, and temperature variations can directly perturb these fiber interfaces, making their polarization stability an important factor in the overall performance of practical quantum communication systems.

Reliable transmission of polarization-encoded quantum states through fiber links requires suppressing fiber-induced polarization drift and relative phase fluctuations. These effects arise mainly from birefringence in the fiber, which can vary with bending, mechanical stress, temperature changes, and other environmental perturbations. Unlike free-space links, where polarization rotations are often dominated by changes in the relative orientation between the transmitter and receiver and can be corrected with relatively simple optical compensation~\cite{takenaka2017satellite, yin2020entanglement}, fiber links generally require compensation of both polarization and phase variation. Active polarization-compensation techniques have therefore been widely studied. Typical implementations rely on optical monitoring, dynamic polarization controllers, high-speed electronic feedback, or wave-plate-based compensation schemes such as combinations of quarter-wave plates (QWP) and half-wave plates (HWP)~\cite{yin2017satellite, simon2012hamilton, tan2024real, yin2025polarization, shi2021fibre}. Measurement-based channel-estimation and compensation methods have also been proposed~\cite{zhou2025efficient, tan2023polarization, luo2024research}. Although these approaches can be effective, they require additional optical and electronic control components, calibration procedures, and feedback algorithms. This can be a practical burden for compact, mobile, or field-deployable quantum communication systems, where simplicity, robustness, and long-term stability are important design requirements. Alternative approaches, such as reference-frame-independent QKD, can reduce sensitivity to reference-frame misalignment~\cite{laing2010reference, lim2025effect}. However, they do not directly remove fiber-induced phase distortions and may require additional measurement resources. These considerations motivate a passive fiber-based stabilization structure that can mitigate fiber-induced polarization and phase fluctuations without relying on real-time active polarization compensation.

In this work, we propose a simple passive stabilization method based on a cross-aligned polarization-maintaining fiber (CAPMF) structure, in which two polarization-maintaining fibers (PMF) are arranged with mutually orthogonal principal axes. Related cross-aligned fiber structures have been used for mode locking in fiber ring lasers and for stabilizing single-photon polarization states, for example in self-compensating polarization encoders~\cite{szczepanek2018nonlinear, agnesi2019all, zhou2019generation}. To the best of our knowledge, however, their applicability to the distribution of nonclassical correlations, particularly polarization-entangled photon pairs, has not been systematically analyzed or experimentally verified. We develop a quantum-information-theoretic model of the CAPMF structure, analyze its entanglement-preserving behavior using fidelity simulations based on the experimental configuration, and experimentally compare SMF and CAPMF fiber connections under environmental perturbations. The results show that the proposed CAPMF structure maintains high interference visibility and preserves the quality of polarization-entangled photon pairs under external perturbations using only a simple passive configuration. This motivates CAPMF as a simple fiber-interface solution for stabilizing polarization entangled state distribution in practical quantum communication systems, including long-distance free-space links~\cite{yin2017satellite, takenaka2017satellite}, drone-assisted QKD~\cite{liu2020drone, lin2020quantum}, and multi-user networks~\cite{chen2021integrated}.

The remainder of this paper is organized as follows. Section~\ref{sec2} presents the CAPMF model and fidelity simulations. Section~\ref{sec3} describes the experimental setup, and Section~\ref{sec4} evaluates the entanglement-preserving performance of the proposed structure. Section~\ref{sec5} concludes the paper.

\section{Theoretical model}\label{sec2}
\subsection{CAPMF transformation model}
Optical fibers are essential components in quantum-optics experiments, where SMFs are commonly used to ensure well-defined spatial-mode propagation and to avoid modal dispersion. However, SMFs are not ideal polarization-preserving media. Environmental variations can cause the fiber birefringence, causing the polarization state propagating through an SMF to undergo time-dependent polarization rotations and relative phase changes~\cite{szczepanek2018nonlinear, agnesi2019all, zhou2019generation}. Consequently, polarization and phase compensation are required for stable polarization-encoded quantum-state transmission.

When the fiber is fixed in a stable laboratory environment, the required compensation parameters may remain nearly constant over time. In practice, stress, vibration, temperature changes, and bending can continuously change the birefringence, making real-time polarization compensation necessary~\cite{tan2024real, yin2025polarization, shi2021fibre, zhou2025efficient, tan2023polarization, luo2024research}. PMFs are widely used to reduce polarization instability by providing well-defined principal axes with strong built-in birefringence. For a fiber of length $L$ at a central wavelength $\lambda$, the phase difference between the two orthogonal polarization axes, $\Delta \varphi$, can be expressed as
\begin{equation}\label{eq1}
\Delta \varphi = \frac{2\pi c}{\lambda}\Delta \tau,
\end{equation}
where $c$ is the speed of light and $\Delta \tau$ denotes the differential group delay (DGD) between the two polarization modes. For a short fiber section with $L<1~\mathrm{km}$, the group delay between the two polarization modes can be approximated as $\Delta \tau=\alpha L$, where $\alpha$ is the polarization-mode-dispersion (PMD) coefficient. At a central wavelength of $\lambda = 810~\mathrm{nm}$, a typical SMF has a PMD coefficient on the order of $\alpha = 10~\mathrm{fs/m}$, whereas a PMF is designed to have a much larger value of approximately $\alpha = 1.2~\mathrm{ps/m}$~\cite{gupta2006polarization, noda2003polarization}. Although Eq.~(\ref{eq1}) applies to both SMFs and PMFs, the large built-in birefringence of PMFs makes the polarization eigenaxes better defined and reduces the relative influence of environmental perturbations compared with SMFs. This stability, however, relies on proper alignment between the input polarization and the PMF eigenaxes. When the input state is not aligned with these axes, the same large birefringence that stabilizes the eigenmodes can introduce a significant relative phase delay between the polarization components of the input state and distort the polarization state.

Let the slow axis of the PMF have a DGD of $\Delta \tau$ relative to the fast axis, and let $\theta$ denote the angular mismatch between the input polarization and the PMF eigenaxes. For a quasi-monochromatic quantum state, the action of the PMF can be described by the Jones operator, $J_\theta(\omega,\Delta \tau)$, as follows~\cite{noda2003polarization, penninckx1999jones}: 
\begin{equation}\label{eq2}
    J_{\theta}(\omega,\Delta\tau)=
    \begin{bmatrix}
        \cos\theta & \sin\theta \\ -\sin \theta & \cos\theta
    \end{bmatrix}
    \begin{bmatrix}
        e^{i\omega\Delta\tau/2} & 0 \\ 0 & e^{-i\omega\Delta\tau/2}
    \end{bmatrix}
    \begin{bmatrix}
        \cos\theta & -\sin\theta \\ \sin \theta & \cos\theta
    \end{bmatrix}.
\end{equation}
Here, $\omega$ is the angular frequency of the photon, and we consider polarization encoding and take the laboratory reference basis to be $\ket{H}=\begin{bmatrix}1, & 0\end{bmatrix}^\mathrm{T}$ and $\ket{V}=\begin{bmatrix}0, & 1\end{bmatrix}^\mathrm{T}$. The PMF eigenaxes can then be related to this basis by the angular mismatch $\theta$ introduced above. In the ideal monochromatic single-mode limit, a PMF can serve as an effective transmission medium for polarization-encoded quantum states, provided that appropriate polarization and phase compensation are applied. In practice, however, quantum light sources have a finite linewidth. The input state, $\ket{\psi_{\mathrm{in}}}$, should therefore be described by including the frequency degree of freedom,
\begin{equation}\label{eq3}
    \ket{\psi_{\mathrm{in}}}=\int d\omega f(\omega)\ket{\omega}\otimes(c_H\ket{H}+c_V\ket{V}).
\end{equation}
Here, $f(\omega)$ denotes the spectral envelope function of the quantum state, and the polarization amplitudes, $c_H$ and $c_v$, satisfy $|c_H|^2+|c_V|^2=1$. After propagation through the PMF described by Eq.~(\ref{eq2}), the output state is given by
\begin{equation}\label{eq4}
    \ket{\psi_{\mathrm{out}}}=\int d\omega f(\omega)\ket{\omega}\otimes J_\theta(\omega,\Delta\tau)(c_H\ket{H}+c_V\ket{V}).
\end{equation}

\begin{figure}
    \centering
    \includegraphics[width=0.8\linewidth]{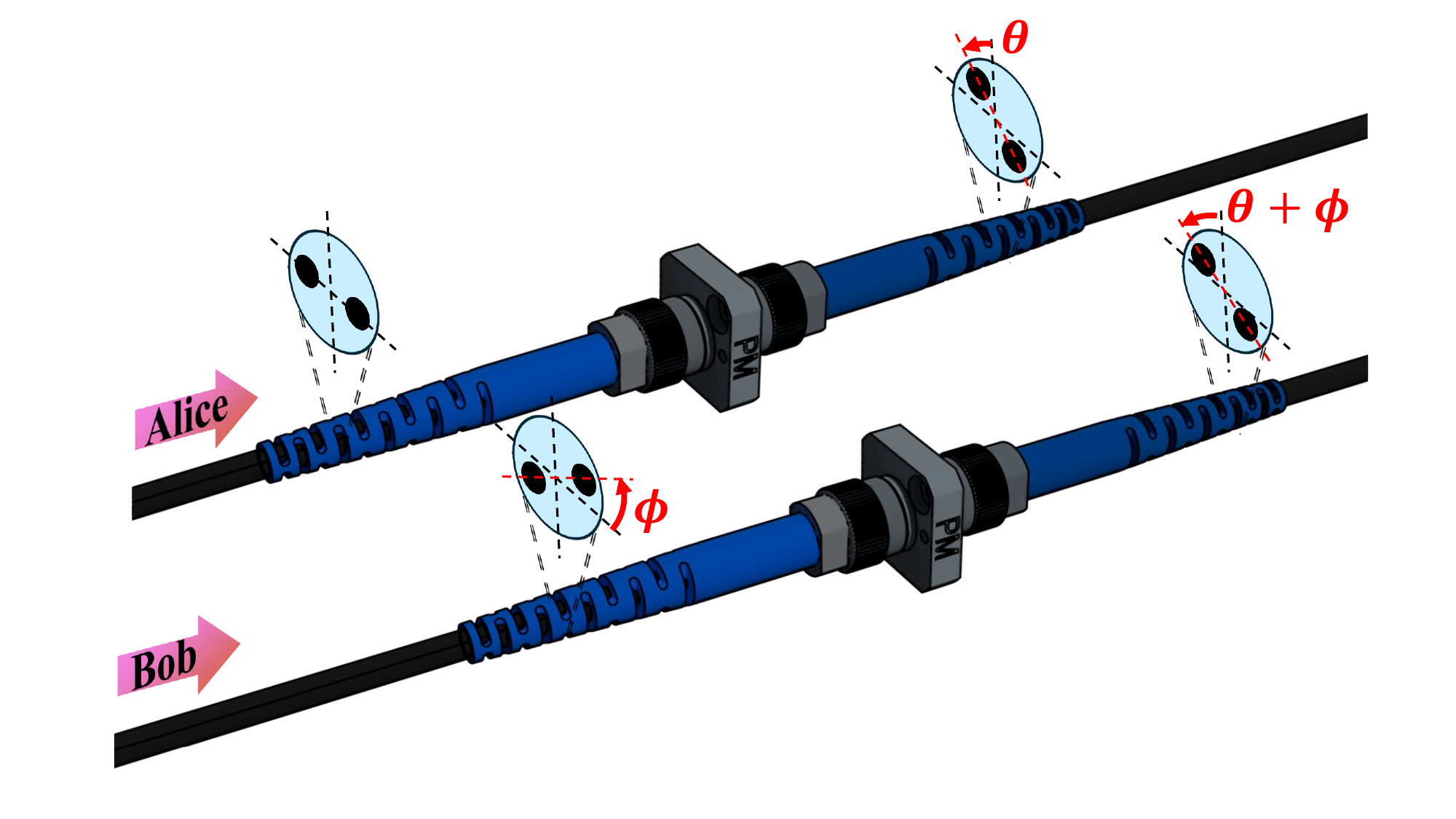}
    \caption{Schematic of the proposed self-compensating transmission system for a bipartite quantum state. Alice and Bob each use a cross-aligned PMF pair. The first PMF on Alice's side defines the horizontal polarization direction of the reference frame. The second PMF on Alice's side is rotated by $90^\circ+\theta$ relative to the first, while the first PMF on Bob's side is rotated by $\phi$ relative to Alice's first PMF. For simple modeling, the first and second PMFs have lengths $L$ and $L+\Delta L$, respectively, for both Alice and Bob.}
    \label{fig:pmf_joint}
\end{figure}
We now apply the single-PMF model to the CAPMF structure, rather than to an individual PMF, as shown in Figure~\ref{fig:pmf_joint}. For the polarization state of Alice and Bob, we define
\[
M_A=J_{\frac{\pi}{2}+\theta}(\omega_A,\Delta\tau')J_0(\omega_A,\Delta\tau), \qquad
M_B=J_{\frac{\pi}{2}+\theta+\phi}(\omega_B,\Delta\tau')J_\phi(\omega_B,\Delta\tau),
\]
respectively, so that the overall transformation is given by $M_A\otimes M_B$.  Here, $\omega_A$ and $\omega_B$ denote the angular frequencies of Alice's and Bob's photons, respectively. The quantities $\Delta\tau$ and $\Delta\tau'$ denote the DGDs of the first and second PMF sections, respectively, and are given by $\Delta\tau=\alpha L$ and $\Delta\tau'=\alpha(L+\Delta L)$. For notational convenience, we write the action of the transformation in the form $M_i\ket{j}=m_i^{(j)}\cdot \vec{x}_i$ as follows:
\begin{equation}\label{eq5}
    m_i^{(j)}=\begin{bmatrix}
        i_1^{(j)} & i_2^{(j)} & i_3^{(j)} & i_4^{(j)}\\
        i_5^{(j)} & i_6^{(j)} & i_7^{(j)} & i_8^{(j)}
    \end{bmatrix},
    \vec{x_i}=\begin{bmatrix}
        e^{i\omega_i(\Delta\tau'+\Delta\tau)/2}\\
        e^{i\omega_i(\Delta\tau'-\Delta\tau)/2}\\
        e^{-i\omega_i(\Delta\tau'-\Delta\tau)/2}\\
        e^{-i\omega_i(\Delta\tau'+\Delta\tau)/2}
    \end{bmatrix}.
\end{equation}
Here, $i\in\{A,B\}$ labels the optical mode associated with each photon of the photon pair, while $j\in\{H,V\}$ labels the input polarization basis state $\ket{H}$ and $\ket{V}$. The explicit forms of $m_A^{(H)}$, $m_A^{(V)}$, $m_B^{(H)}$, and $m_B^{(V)}$ corresponding to $M_A$ and $M_B$ are given in Appendix~\ref{secA1}. With this notation, the transformation of a monochromatic $\ket{\psi^-}$ Bell state through the CAPMF pair can be written as
\begin{equation}\label{eq6}
    M_A\otimes M_B \cdot \frac{\ket{HV}_{AB}-\ket{VH}_{AB}}{\sqrt{2}}
    = \left(m_A^{(H)}\otimes m_B^{(V)}-m_A^{(V)}\otimes m_B^{(H)} \right)\cdot(\vec{x}_A\otimes \vec{x}_B).
\end{equation}

\subsection{Entanglement fidelity through CAPMF}
We now extend the polarization-transformation model to SPDC entangled photon pairs. Polarization-entangled photon pairs are commonly generated by spontaneous parametric down-conversion (SPDC), for which the state can be written as~\cite{steinlechner2014efficient, grice1997spectral, u2006generation}
\begin{equation}\label{eq7}
\ket{\psi_{\mathrm{SPDC}}}=
\int\int d\omega_A d\omega_B
e^{i\Phi(\omega_A,\omega_B)}\sqrt{S(\omega_A,\omega_B)}
\ket{\omega_A,\omega_B}\otimes\ket{\psi^-}.
\end{equation}
Here, $S(\omega_A,\omega_B)$ denotes the joint spectral intensity, and $\Phi(\omega_A,\omega_B)$ is the joint spectral phase. The effects of the pump spectrum and phase matching are absorbed into the effective joint spectral intensity and joint spectral phase. For simplicity, we adopt the perfect phase-matching approximation, $\Phi(\omega_A,\omega_B)=0$, which is consistent with operating the PPKTP crystal near its phase-matching condition. We then model the effective joint spectral intensity $S(\omega_A,\omega_B)$ as a bivariate Gaussian function with correlated signal and idler frequencies:
\begin{equation}\label{eq8}
S(\omega_A,\omega_B)=K_1e^{K_2\left(x^2-2\kappa xy + y^2\right)},
\end{equation}
where
$K_1=\frac{1}{2\pi\sigma_\omega^2\sqrt{1-\kappa^2}}$,
$K_2=-\frac{1}{2(1-\kappa^2)}$,
$x=\frac{\omega_A-\omega_0}{\sigma_\omega}$, and
$y=\frac{\omega_B-\omega_0}{\sigma_\omega}$.
Here, $\omega_0$ is the central angular frequency, and $\sigma_\omega$ is the marginal angular-frequency standard deviation. The parameter $\kappa$ denotes the frequency correlation between $\omega_A$ and $\omega_B$, defined as $\kappa=E[x\cdot y]$.

By extending the state-transformation formalism in Eq.~(\ref{eq4}), we calculate the output state after the entangled state in Eq.~(\ref{eq7}) passes through the CAPMF structures of Alice and Bob. The frequency-dependent phase products generated by the CAPMF transformation are incorporated through a spectral coherence matrix obtained by averaging over $S(\omega_A,\omega_B)$. The reduced polarization density matrix, $\rho_o$, after the CAPMF transformation and frequency-unresolved measurement is written as
\begin{equation}\label{eq9}
\rho_o=\int\int d\omega_A d\omega_B S(\omega_A,\omega_B)
\ket{\chi(\omega_A,\omega_B)}\bra{\chi(\omega_A,\omega_B)},
\end{equation}
where $\ket{\chi(\omega_A,\omega_B)}=[M_A(\omega_A)\otimes M_B(\omega_B)]\ket{\psi^-}$, with $M_A$ and $M_B$ defined in the same form as Eq.~(\ref{eq6}). A detailed derivation, including the explicit form of the resulting density matrix, is provided in Appendix~\ref{secA2}. The fidelity is calculated directly from the frequency-traced polarization density matrix as
\begin{equation}\label{eq10}
F=\bra{\psi^-}\rho_{\mathrm{o}}\ket{\psi^-}.
\end{equation}

\begin{figure}[t]
\centering
\includegraphics[width=0.87\linewidth]{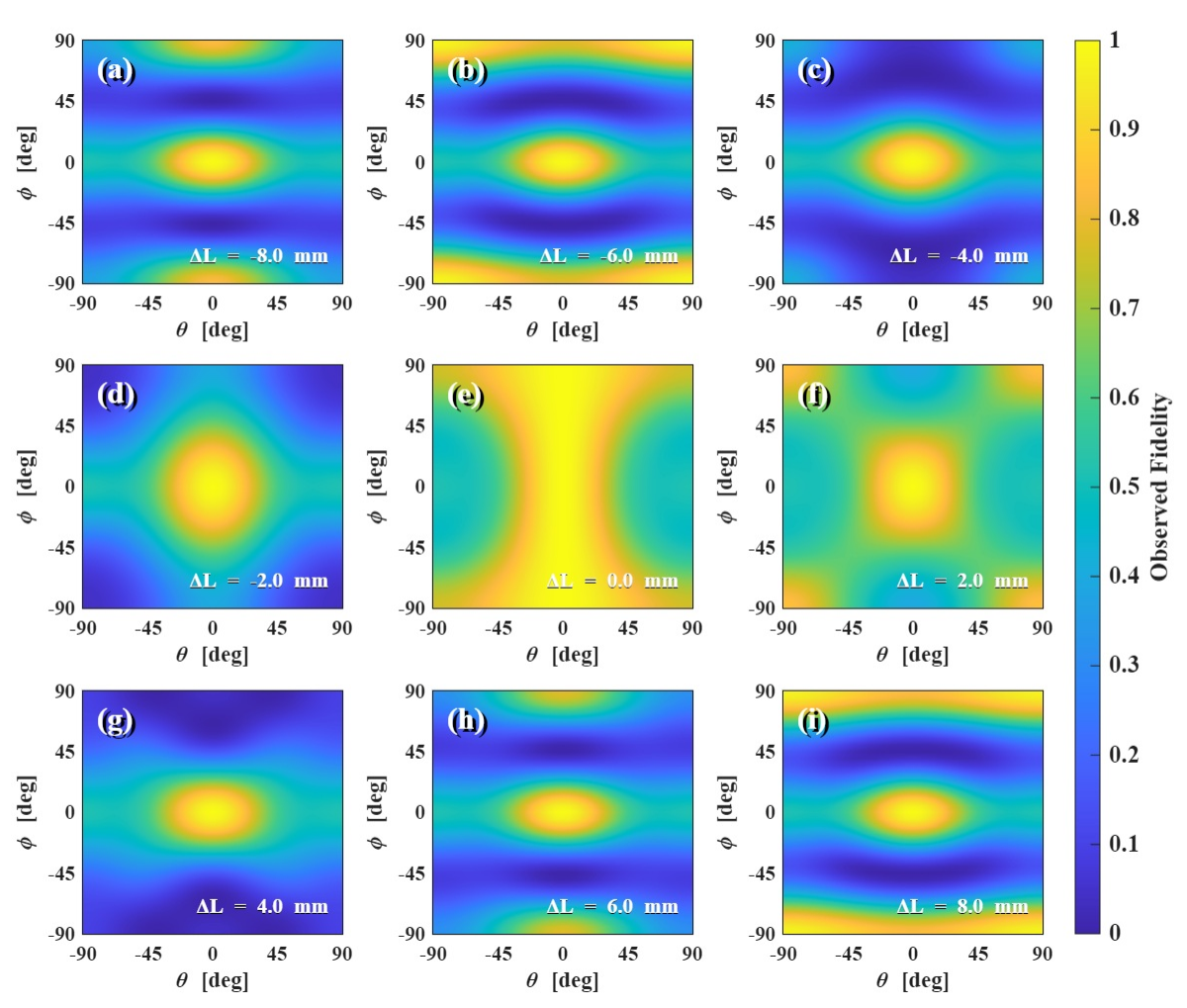}
\caption{
Simulated fidelity for transmitting a $\ket{\psi^-}$ photon pair at $\lambda=811.6~\mathrm{nm}$ through cross-aligned PMF pairs. The reference PMF length is $L=0.7~\mathrm{m}$, the PMF DGD coefficient is $\alpha=1.16~\mathrm{ps/m}$. The parameters $\theta$ and $\phi$ denote the PMF fast-axis alignment error and the Alice--Bob reference-frame mismatch, respectively. Each subplot (a)--(i) corresponds to a different length mismatch $\Delta L$ from $-8$ to $8~\mathrm{mm}$, as indicated in the panel. 
}
\label{fig:pmf_sim}
\end{figure}

Using the model defined above, we calculate $F$ as a function of the alignment error $\theta$ and the length mismatch $\Delta L$, as shown in Figure~\ref{fig:pmf_sim}. In the simulation, the marginal angular-frequency standard deviation is set to $\sigma_\omega=8.0\times10^{11}~\mathrm{s}^{-1}$, which is estimated from the single-photon coherence time of the signal and idler photons. The spectral correlation coefficient is set to $\kappa=-0.99$ to represent the strong frequency anticorrelation of the CW-pumped SPDC photon pairs. These values are used to represent the finite joint spectral bandwidth of the detected SPDC photon pairs under our experimental condition.

High fidelity is maintained when both $\theta$ and $|\Delta L|$ are small, and it decreases gradually rather than abruptly as these parameters increase. This indicates that the CAPMF structure has a finite tolerance to practical alignment and length errors. Near the ideal cross-alignment condition, the residual DGD remains sufficiently small to preserve the compensation, while moderate deviations do not immediately destroy the compensated transmission. The simulation shows that the fidelity is determined by the combined effects of the CAPMF alignment error $\theta$ and the relative polarization-frame mismatch $\phi$. The parameter $\theta$ changes the internal cross alignment of the two PMF sections and therefore modifies how the fast- and slow-axis components are recombined after propagation. In contrast, $\phi$ changes the relative polarization basis between Alice and Bob and shifts the interference condition of the output polarization components. As a result, deviations in either parameter can reduce the overlap with the target $\ket{\psi^-}$ state, but the compensation remains stable within a finite region around the ideal alignment condition. This behavior indicates that CAPMF does not require perfect alignment, but instead provides a practical tolerance window for maintaining high-fidelity entanglement distribution.

Under the present simulation conditions, using $F>0.95$ as the criterion for high fidelity, stable compensation performance is maintained over the ranges $|\Delta L|\le 8~\mathrm{mm}$, $|\theta|\le 5^\circ$, and $|\phi|\le 5^\circ$. These results suggest that CAPMF provides passive stabilization with experimentally accessible alignment requirements.

\section{Experimental scheme}\label{sec3}
\begin{figure}
    \centering
    \includegraphics[width=0.8\linewidth]{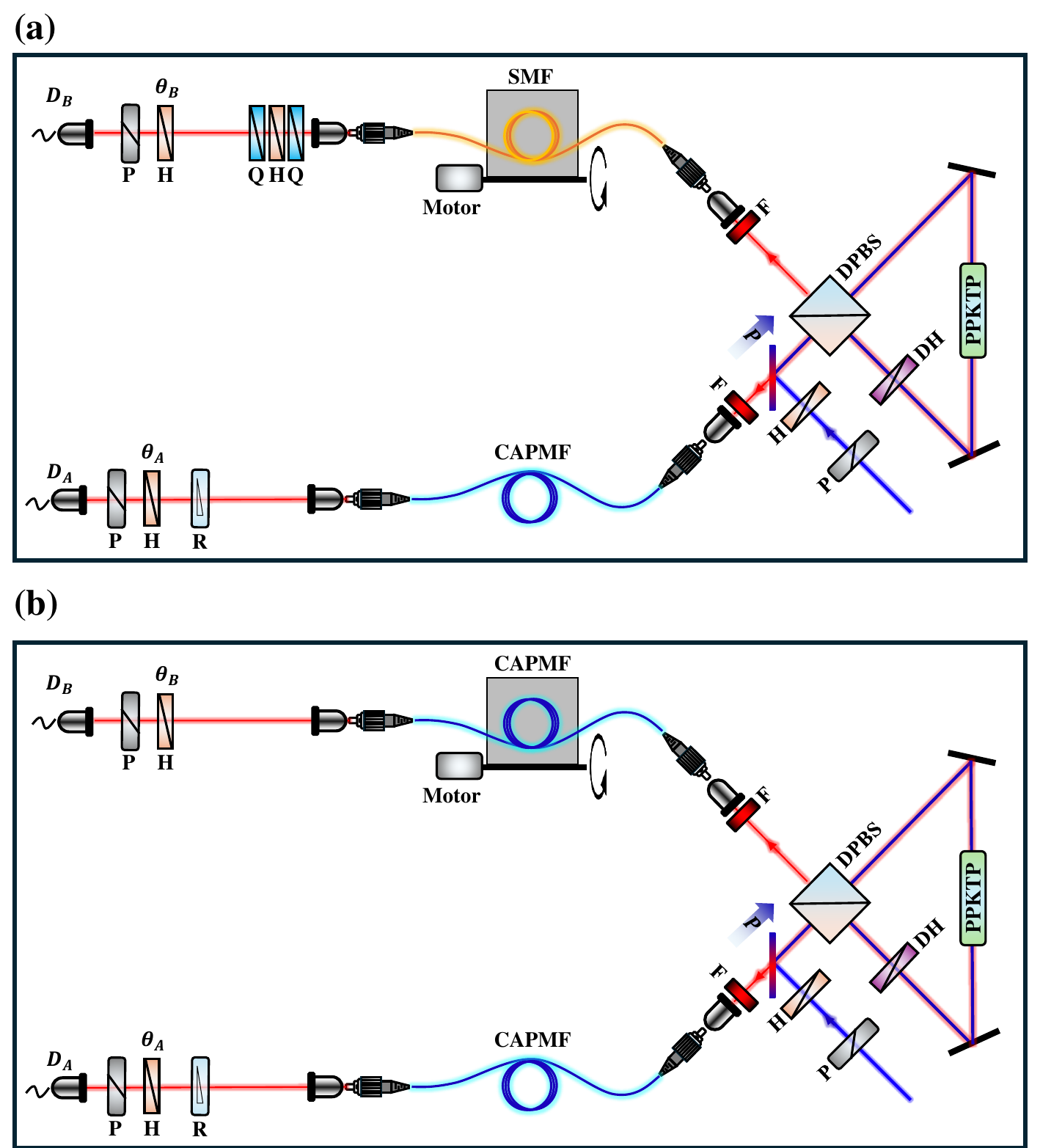}
    \caption{
    Experimental schematic for entangled-photon-pair generation and measurement. The entangled photon pairs are generated in a Sagnac interferometer, and the measurement basis is selected by linear polarizers in the measurement module. Alice and Bob detect the photons with SPADs, while a TCSPC module records single-count and coincidence-count data. (a) SMF configuration, in which the source and measurement module are connected through an SMF interface and one HWP and two QWPs are used for polarization and phase compensation. (b) CAPMF configuration, in which the source and measurement module are connected through a CAPMF pair without additional polarization-compensation wave plates. H: half-wave plate; Q: quarter-wave plate; R: phase retarder; PPKTP: periodically poled potassium titanyl phosphate; DPBS: dual-wavelength polarizing beam splitter; DH: dual-wavelength half-wave plate; F: bandpass filter;  P: linear polarizer.
    }
    \label{fig:exp_scheme}
\end{figure}

The experimental setup is shown in Figure~\ref{fig:exp_scheme}. It consists of three main parts: a Sagnac interferometer for generating polarization-entangled photon pairs, a measurement module for detecting the distributed entangled photons, and an interfacing fiber that connects the two independent optical systems. The Sagnac interferometer generates degenerate entangled photon pairs through spontaneous parametric down-conversion (SPDC) using a type-II PPKTP crystal and a pump laser with a central wavelength of $405.8~\mathrm{nm}$. The PPKTP crystal has a length of $20~\mathrm{mm}$ and a poling period of $10~\mu\mathrm{m}$, and it is heated to $80 \pm 0.1^{\circ}\mathrm{C}$ to satisfy the phase-matching condition at the pump wavelength~\cite{fan1987second, konig2004extended, Wiechmann1993}.
The source uses a Sagnac interferometer to overlap the clockwise (CW) and counterclockwise (CCW) SPDC modes, preparing polarization-entangled photon pairs in the Bell state $\ket{\psi^-}$~\cite{Kim2024HighBrightnessCHSH, steinlechner2014efficient}. A $10~\mathrm{nm}$ bandpass filter centered at $810~\mathrm{nm}$ is used to spectrally select the down-converted photons and suppress spectral distinguishability between the photon-pair components. At a fixed pump power of $P=3~\mathrm{mW}$, the photon-pair generation rate is $110.7~\mathrm{kcps/mW}$. The heralding efficiency is $\eta=N_c/\sqrt{N_A N_B}=20.2\%$, where $N_A$ and $N_B$ are the single counts of Alice and Bob, and $N_c$ is the coincidence count.

The measurement module for the distributed entangled photon pairs consists of a linear polarizer and a single-photon avalanche diode (SPAD) on each side, enabling Alice and Bob to choose the polarization measurement basis. The detectors are denoted by $D_i$ with $i\in\{A,B\}$. The SPADs have a detection efficiency of approximately $60\%$ near $810~\mathrm{nm}$, a typical dark-count rate of about $500~\mathrm{cps}$ per channel, and a dead time of $50~\mathrm{ns}$. The SPAD signals are recorded using a time-correlated single-photon counting (TCSPC) module with picosecond timing resolution, from which both single counts and coincidence counts are obtained. The coincidence window is set to $T_{\mathrm{cw}}=2\,\mathrm{ns}$ for identifying temporally correlated photon-pair events. With this measurement setting, we compare the two interfacing configurations shown in Figure ~\ref{fig:exp_scheme}.

For the SMF configuration, polarization compensation is required because bending of the fiber during the initial setup can introduce polarization and phase variations that affect accurate entanglement-state measurements. To compensate for these variations, two quarter-wave plates (QWPs) and one half-wave plate (HWP) are placed on each of the Alice and Bob paths. During the experiment, the laboratory temperature and humidity are kept stable to minimize time-dependent polarization drift. Under these conditions, the fixed polarization-compensation optics remain stable during the measurement, provided that the SMF is not physically perturbed.

In contrast, for the CAPMF configuration, the polarization state is maintained by the fiber structure itself, and the wave plates used for polarization compensation are omitted. In both the SMF and CAPMF configurations, the relative phase between Alice and Bob is adjusted to prepare the $\ket{\psi^-}$ state using a horizontally aligned QWP tilted to act as a fixed phase retarder. The two configurations use the same source and measurement modules, while only the interfacing fiber between them is replaced by either an SMF or a CAPMF. Because this replacement requires realignment, small differences in optical alignment, loss, and visibility can occur between the measurements. To minimize such effects, the optical alignment is repeated before each measurement, and thus these small differences do not affect the main purpose or conclusion of the experiment.

The experiment is performed by measuring coincidence-count interference fringes under four conditions: SMF or CAPMF as the interfacing fiber, and fixed or externally perturbed fiber conditions. For the CAPMF configuration, two $0.7~\mathrm{m}$ PMFs with a length mismatch within approximately $\pm 2~\mathrm{mm}$ are precisely spliced with mutually orthogonal principal axes. For the perturbation test, both the $1.4~\mathrm{m}$ SMF and CAPMF samples are fixed in the same circular shape with a diameter of approximately $10~\mathrm{cm}$ and mounted on a rigid plate. In this way, the two configurations experience the same bending diameter and motor-driven rotation condition. In the CAPMF case, this circular arrangement also allows the two PMF sections to experience similar bending perturbations, which helps maintain the temporal-walk-off compensation condition. The perturbation is applied by rotating the fiber-mounted plate over a $90^\circ$ range at $1~\mathrm{Hz}$. We refer to the fixed fiber condition as the ``stable'' condition and the motor-driven bending condition as the ``unstable'' condition.

\section{Experimental Results}\label{sec4}
\begin{figure}[t]
    \centering
    \includegraphics[width=0.9\linewidth]{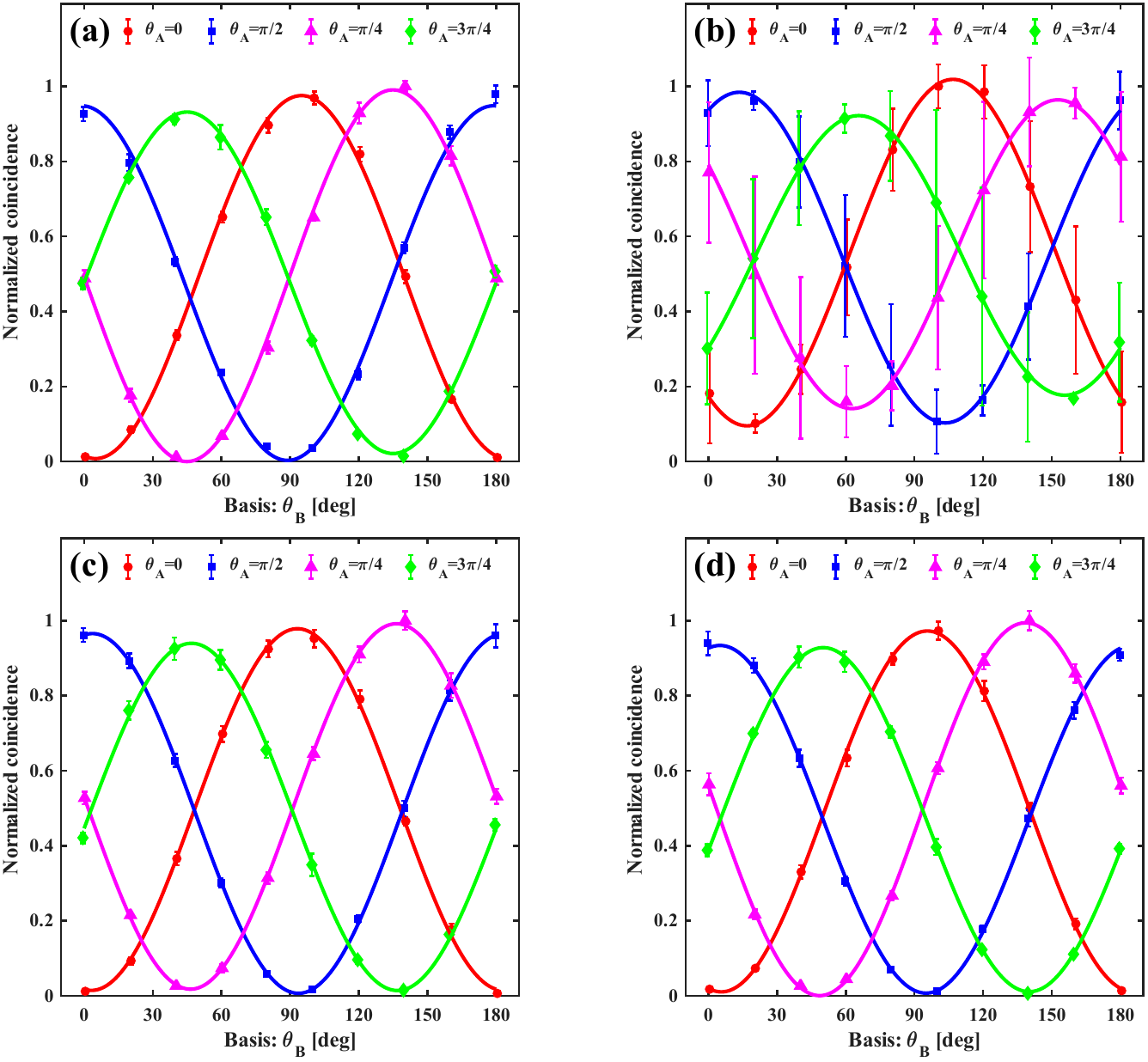}
    \caption{
    Interference fringes of entangled photon pairs measured using an SMF and a cross-aligned PMF pair. (a) and (b) correspond to the SMF case under stable and unstable conditions, respectively. (c) and (d) correspond to the CAPMF case under stable and unstable conditions, respectively. Each fringe is obtained by fixing Alice's measurement basis and scanning Bob's measurement angle $\theta_B$, while recording coincidence counts between $D_A$ and $D_B$. The CAPMF pair preserves high visibility even under external bending and mechanical perturbations.
    }
    \label{fig:exp_result}
\end{figure}

With Alice's measurement basis fixed, we measure coincidence-count interference fringes by scanning Bob's polarizer angle $\theta_B$. CHSH measurements are also performed for each configuration to verify the preservation of nonlocal correlations. Figure~\ref{fig:exp_result} shows the measured fringes for the SMF and CAPMF configurations under stable and unstable conditions. Figure~\ref{fig:exp_result}(a) and Figure~\ref{fig:exp_result}(b) correspond to the SMF case, while Figure~\ref{fig:exp_result}(c) and Figure~\ref{fig:exp_result}(d) correspond to the CAPMF case. The stable condition is shown in Figure~\ref{fig:exp_result}(a) and Figure~\ref{fig:exp_result}(c), and the unstable condition with external bending and mechanical perturbation is shown in Figure~\ref{fig:exp_result}(b) and Figure~\ref{fig:exp_result}(d). Each curve is obtained by fixing Alice's projection basis to one of four polarization states and scanning Bob's measurement angle $\theta_B$ from $0^\circ$ to $180^\circ$. Specifically, $\theta_A=0^\circ$ and $90^\circ$ correspond to projections onto $\ket{H}$ and $\ket{V}$, respectively, while $\theta_A=45^\circ$ and $135^\circ$ correspond to projections onto $\ket{D}$ and $\ket{A}$, respectively.

Replacing the interfacing fiber can introduce small variations in optical coupling efficiency and total count rate between experimental conditions. Therefore, the coincidence counts in Figure~\ref{fig:exp_result} are normalized by the maximum coincidence count of each dataset, which is listed as the normalization factor in Table~\ref{tab:statistics}. Since the visibility and CHSH value are invariant under a common scaling of the coincidence counts, these count-rate variations do not affect the main conclusions of this work.

Table~\ref{tab:statistics} summarizes the main statistical quantities measured under stable and unstable conditions for the SMF and CAPMF configurations. The normalization factor is the maximum coincidence count used for fringe normalization. The visibility is defined as $V=(C_{\max}-C_{\min})/(C_{\max}+C_{\min})$, where $C_{\max}$ and $C_{\min}$ are the maximum and minimum coincidence counts of the fitted fringe, respectively. For the SMF case, the stable condition yields a high average visibility of $0.9830\pm0.0033$, and the CHSH value is $S=2.7852\pm0.0234$, clearly exceeding the classical bound of 2. This shows that, in a stable laboratory environment without external perturbations, an SMF can maintain high interference contrast and nonlocal correlations when used with a fixed polarization-compensation system after initial alignment. However, the performance of the same SMF configuration degrades substantially under the unstable condition. The average visibility decreases to $0.7655\pm0.0523$, and the CHSH value drops to $S=1.7714\pm0.2779$, so that Bell-inequality violation is no longer maintained. The uncertainty of the visibility also increases significantly compared with the stable condition, indicating that external bending and mechanical perturbations induce time-dependent polarization fluctuations inside the SMF.

\begin{table}[t]
\centering
\caption{Fringe statistics for SMF and CAPMF configurations under stable and unstable conditions.}
\begin{tabular}{lcccc}
\toprule
\textbf{Metric} & \multicolumn{2}{c}{\textbf{SMF configuration}} & \multicolumn{2}{c}{\textbf{CAPMF configuration}} \\
\cmidrule(lr){2-3} \cmidrule(lr){4-5}
                   & (a) stable & (b) unstable & (c) stable & (d) unstable \\ 
\midrule
Normalization factor [cps]                     
& 2056.3 & 2031.8 & 1883.4 & 1832.0 \\
Visibility                       
& 0.9830$\pm$0.0033  & 0.7655$\pm$0.0523 & 0.9727$\pm$0.0029  & 0.9843$\pm$0.0036 \\
CHSH value                       
& 2.7852$\pm$0.0234 & 1.7714$\pm$0.2779 & 2.7764$\pm$0.0258 & 2.6838$\pm$0.0291 \\
\bottomrule
\end{tabular}
\label{tab:statistics}
\end{table}

\begin{figure}[t]
    \centering
    \includegraphics[width=0.9\linewidth]{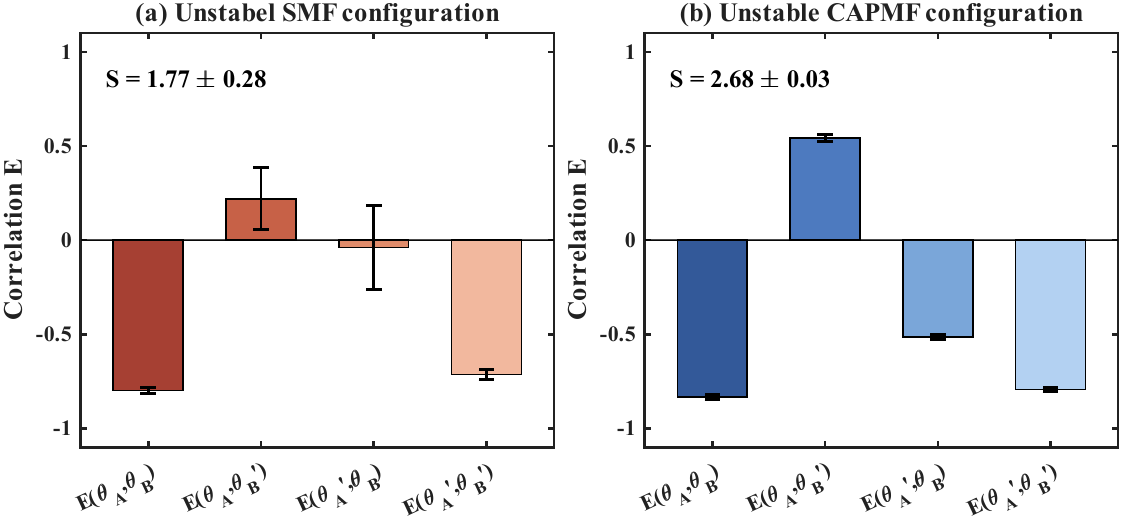}
    \caption{
    CHSH correlation results measured under the unstable SMF and CAPMF configurations.
    }
    \label{fig:chsh_result}
\end{figure}
In contrast, the CAPMF pair exhibits much more stable behavior under external perturbations. Under the stable condition, it shows a high average visibility of $0.9727\pm0.0029$ and a CHSH value of $S=2.7764\pm0.0258$, comparable to the compensated SMF configuration. More importantly, under the unstable condition, the CAPMF configuration maintains an average visibility of $0.9843\pm0.0036$ and a CHSH value of $S=2.6838\pm0.0291$, clearly preserving Bell-inequality violation.

Figure~\ref{fig:chsh_result} compares the four polarization correlation coefficients measured under the unstable SMF and CAPMF configurations. The CHSH value, $S$, is calculated using the measurement settings $\theta_A=0^\circ$, $\theta_A'=45^\circ$, $\theta_B=22.5^\circ$, and $\theta_B'=67.5^\circ$ as~\cite{clauser1969proposed}
\begin{equation}\label{eq11}
    S =
    \left|
    E(\theta_A,\theta_B)
    -
    E(\theta_A,\theta_B')
    +
    E(\theta_A',\theta_B)
    +
    E(\theta_A',\theta_B')
    \right|,
\end{equation}
where $E(\theta_A,\theta_B)$ is the polarization correlation coefficient for the corresponding pair of measurement bases $\theta_A$ and $\theta_B$. In the unstable SMF case, the four correlation coefficients are strongly degraded, whereas in the unstable CAPMF case they remain well preserved across the required measurement settings. This contrast is consistent with the corresponding CHSH results summarized in Table~\ref{tab:statistics}, namely the loss of Bell-inequality violation for SMF and its preservation for CAPMF. These results indicate that the CAPMF pair preserves not only fringe visibility but also the nonlocal correlations of the entangled photon pairs under mechanical perturbations.

Overall, the SMF configuration is strongly affected by external perturbations, showing substantial degradation in both visibility and CHSH correlation. In contrast, the CAPMF pair maintains high visibility and a clear Bell-inequality violation even under the unstable condition. These results experimentally demonstrate that CAPMF can provide both polarization stability and entanglement preservation required for field-operated quantum communication systems. The compensation effect, however, assumes that the two PMF sections experience similar perturbations; strong asymmetric stress or temperature gradients between the sections may therefore require further design optimization. For example, repeated short CAPMF sections may help reduce sensitivity to spatially nonuniform perturbations and preserve polarization entanglement during fiber-interface transmission~\cite{szczepanek2018nonlinear}.

\section{Conclusions}\label{sec5}
In this work, we theoretically analyzed and experimentally verified stable entanglement \textcolor{blue}{distribution} using a CAPMF structure, in which two PMFs are arranged with mutually orthogonal principal axes. The CAPMF structure compensates the polarization transformation and temporal walk-off generated in the two PMF sections, thereby preserving the polarization correlations of entangled photon pairs without real-time active polarization compensation. Experimentally, the SMF configuration exhibited high visibility and a clear CHSH violation under the stable condition. Under the unstable condition, however, its average visibility decreased to $0.7655\pm0.0523$, and the CHSH value dropped to $S=1.7714\pm0.2779$, so that Bell-inequality violation was no longer maintained. In contrast, the CAPMF pair maintained an average visibility of $0.9843\pm0.0036$ and a CHSH value of $S=2.6838\pm0.0291$ under the same unstable condition. These results show that CAPMF can effectively suppress fiber-induced polarization fluctuations and preserve entanglement quality under external mechanical perturbations. Together with the theoretical model, the experimental results support the role of cross-aligned PMF sections in compensating temporal walk-off. Overall, the CAPMF pair provides a passive entanglement-transmission structure that maintains high visibility, low uncertainty, and stable CHSH violation without a complex active polarization-compensation system. These results demonstrate that CAPMF can passively stabilize fiber-interface connecting photon sources to transmitting/receiving optics and measurement modules while preserving polarization entanglement under mechanical perturbations. This makes CAPMF a practical candidate for quantum communication systems where short fiber interfaces are unavoidable. Extending the evaluation to broader operating conditions, such as different fiber lengths, temperature variations, bending conditions, vibration spectra, and long-term operation, will further support the development of CAPMF-based interfaces for field-deployable quantum communication systems.

\begin{appendices}




\section{CAPMF coefficient matrices}\label{secA1}
The CAPMF operators for Alice and Bob are given by
$M_A=J_{\frac{\pi}{2}+\theta}(\omega_A,\Delta\tau')J_0(\omega_A,\Delta\tau)$ and
$M_B=J_{\frac{\pi}{2}+\theta+\phi}(\omega_B,\Delta\tau')J_\phi(\omega_B,\Delta\tau)$, respectively.
The action of each operator on the polarization basis states
$\ket{H}\equiv\ket{0}$ and $\ket{V}\equiv\ket{1}$ can be expressed as the product of a $2\times4$ coefficient matrix $m_i^{(j)}$ and a vector $\vec{x}_i$ containing four phase factors:
\begin{equation}
    M_i\ket{j}=m_i^{(j)}\vec{x}_i,
    \qquad
    i\in\{A,B\},\quad j\in\{H,V\},
\end{equation}
where
\begin{equation}
    m_i^{(j)}=
    \begin{bmatrix}
        i_1^{(j)} & i_2^{(j)} & i_3^{(j)} & i_4^{(j)}\\
        i_5^{(j)} & i_6^{(j)} & i_7^{(j)} & i_8^{(j)}
    \end{bmatrix}.
\end{equation}
Here, the superscript $j$ denotes the input polarization state, while the subscript indexes the coefficient corresponding to each phase factor. Since the fast axis of Alice's first PMF is chosen as the $\ket{H}$ direction of the reference frame, the coefficients of $m_A^{(H)}$ and $m_A^{(V)}$ are simplified as
\begin{equation}
\begin{aligned}
&A_1^{(H)}=\sin^2\theta,          &&A_2^{(H)}=0,
&&A_3^{(H)}=\cos^2\theta,          &&A_4^{(H)}=0,\\
&A_5^{(H)}=\sin\theta\cos\theta,  &&A_6^{(H)}=0,
&&A_7^{(H)}=-\sin\theta\cos\theta, &&A_8^{(H)}=0,
\end{aligned}
\end{equation}
and
\begin{equation}
\begin{aligned}
&A_1^{(V)}=0,                     &&A_2^{(V)}=\sin\theta\cos\theta,
&&A_3^{(V)}=0,                     &&A_4^{(V)}=-\sin\theta\cos\theta,\\
&A_5^{(V)}=0,                     &&A_6^{(V)}=\cos^2\theta,
&&A_7^{(V)}=0,                     &&A_8^{(V)}=\sin^2\theta.
\end{aligned}
\end{equation}
Therefore, the coefficient matrices for Alice's arm can be written as
\begin{equation}
    m_A^{(H)}
    =
    \begin{bmatrix}
        \sin^2\theta & 0 & \cos^2\theta & 0\\
        \sin\theta\cos\theta & 0 & -\sin\theta\cos\theta & 0
    \end{bmatrix},
\end{equation}
\begin{equation}
    m_A^{(V)}
    =
    \begin{bmatrix}
        0 & \sin\theta\cos\theta & 0 & -\sin\theta\cos\theta\\
        0 & \cos^2\theta & 0 & \sin^2\theta
    \end{bmatrix}.
\end{equation}
For Bob's arm, the reference-frame difference $\phi$ between Alice and Bob is included, and the coefficients are given by
\begin{equation}
\begin{aligned}
&B_1^{(H)}=\cos\phi\sin\theta\sin(\theta+\phi),
&&B_2^{(H)}=\sin\phi\cos\theta\sin(\theta+\phi),\\
&B_3^{(H)}=\cos\phi\cos\theta\cos(\theta+\phi),
&&B_4^{(H)}=-\sin\phi\sin\theta\cos(\theta+\phi),\\
&B_5^{(H)}=\cos\phi\sin\theta\cos(\theta+\phi),
&&B_6^{(H)}=\sin\phi\cos\theta\cos(\theta+\phi),\\
&B_7^{(H)}=-\cos\phi\cos\theta\sin(\theta+\phi),
&&B_8^{(H)}=\sin\phi\sin\theta\sin(\theta+\phi),
\end{aligned}
\end{equation}
and
\begin{equation}
\begin{aligned}
&B_1^{(V)}=-\sin\phi\sin\theta\sin(\theta+\phi),
&&B_2^{(V)}=\cos\phi\cos\theta\sin(\theta+\phi),\\
&B_3^{(V)}=-\sin\phi\cos\theta\cos(\theta+\phi),
&&B_4^{(V)}=-\cos\phi\sin\theta\cos(\theta+\phi),\\
&B_5^{(V)}=-\sin\phi\sin\theta\cos(\theta+\phi),
&&B_6^{(V)}=\cos\phi\cos\theta\cos(\theta+\phi),\\
&B_7^{(V)}=\sin\phi\cos\theta\sin(\theta+\phi),
&&B_8^{(V)}=\cos\phi\sin\theta\sin(\theta+\phi).
\end{aligned}
\end{equation}
Thus, the coefficient matrices for Bob's arm are
\begin{equation}
    m_B^{(H)}
    =
    \begin{bmatrix}
        B_1^{(H)} & B_2^{(H)} & B_3^{(H)} & B_4^{(H)}\\
        B_5^{(H)} & B_6^{(H)} & B_7^{(H)} & B_8^{(H)}
    \end{bmatrix},
\end{equation}
\begin{equation}
    m_B^{(V)}
    =
    \begin{bmatrix}
        B_1^{(V)} & B_2^{(V)} & B_3^{(V)} & B_4^{(V)}\\
        B_5^{(V)} & B_6^{(V)} & B_7^{(V)} & B_8^{(V)}
    \end{bmatrix}.
\end{equation}
Using these matrix elements, the polarization state after the $\ket{\psi^-}$ Bell state passes through Alice's and Bob's CAPMF structures can be calculated, as in Eq.~(\ref{eq6}), in the form
\[
\left(m_A^{(H)}\otimes m_B^{(V)}-m_A^{(V)}\otimes m_B^{(H)}\right)
(\vec{x}_A\otimes\vec{x}_B).
\]

\section{Evaluation of the joint spectral integral}\label{secA2}
We define
$K_1=\frac{1}{2\pi\sigma_\omega^2\sqrt{1-\kappa^2}}$,
$K_2=-\frac{1}{2(1-\kappa^2)}$,
$x=\frac{\omega_A-\omega_0}{\sigma_\omega}$, and
$y=\frac{\omega_B-\omega_0}{\sigma_\omega}$.
The joint spectral envelope
$S(\omega_A,\omega_B)=K_1e^{K_2(x^2-2\kappa xy+y^2)}$
satisfies the normalization condition
\begin{equation}\label{eqB3}
    \int\int d\omega_A d\omega_B S(\omega_A,\omega_B)=1.
\end{equation}
Under the change of variables, the integration measure becomes
$d\omega_A d\omega_B = \sigma_\omega^2 dxdy$.
Separating the central frequency contribution gives
\begin{equation}\label{eqB4}
    e^{i\omega_A t_A}e^{i\omega_B t_B}
    =
    e^{i\omega_0(t_A+t_B)}e^{i\sigma_\omega(xt_A+yt_B)}.
\end{equation}
Thus, $I(t_A,t_B)$ can be written as
\begin{equation}\label{eqB5}
    I(t_A,t_B)=
    \frac{e^{i\omega_0 (t_A+t_B)}}{2\pi\sqrt{1-\kappa^2}}
    \int\int dxdy \;
    \exp\left[
    -\frac{x^2-2\kappa xy+y^2}{2(1-\kappa^2)}
    +i\sigma_\omega(xt_A+yt_B)
    \right].
\end{equation}
The exponent can be expressed in matrix form as
\begin{equation}\label{eqB6}
    -\frac{x^2-2\kappa xy+y^2}{2(1-\kappa^2)}
    +i\sigma_\omega(xt_A+yt_B)
    =
    -\frac{1}{2}\vec{X}^{T}\Sigma^{-1}\vec{X}
    +i\sigma_\omega \vec{T}^{T}\vec{X},
\end{equation}
where
\[
\vec{X}=
\begin{bmatrix}
x \\ y
\end{bmatrix},
\qquad
\vec{T}=
\begin{bmatrix}
t_A \\ t_B
\end{bmatrix},
\qquad
\Sigma=
\begin{bmatrix}
1 & \kappa \\
\kappa & 1
\end{bmatrix}.
\]
Using the standard Gaussian integral
\begin{equation}
    \int d^n x\,
    e^{-\frac{1}{2}x^T A x+b^T x}
    =
    \sqrt{\frac{(2\pi)^n}{\det A}}\,
    e^{\frac{1}{2}b^T A^{-1}b},
\end{equation}
with $A=\Sigma^{-1}$ and $b=i\sigma_\omega\vec{T}$, Eq.~(\ref{eqB5}) reduces to
\begin{equation}
    I(t_A,t_B)=
    e^{i\omega_0(t_A+t_B)}
    e^{-\frac{\sigma_\omega^2}{2}(t_A^2+t_B^2+2\kappa t_At_B)}.
\end{equation}

The result above is used to evaluate the spectral coherence terms that appear after the CAPMF transformation. In Eq.~(\ref{eq9}), the operators $M_A$ and $M_B$ contain products of phase factors associated with Alice's and Bob's frequencies. In the reduced polarization density matrix, these phase factors appear as products between the ket and bra components. Each such term can be written in the form
\begin{equation}
e^{i\omega_A \Delta t_A}e^{i\omega_B \Delta t_B},
\end{equation}
where $\Delta t_A$ and $\Delta t_B$ are the differences between the corresponding DGD-induced time delays in the ket and bra components. Its spectral average over the joint spectral intensity is then given by
\begin{equation}
\int\int d\omega_A d\omega_B,
S(\omega_A,\omega_B)
e^{i\omega_A \Delta t_A}
e^{i\omega_B \Delta t_B}
=
I(\Delta t_A,\Delta t_B).
\end{equation}
Therefore, the finite spectral bandwidth of the SPDC photons is incorporated through the spectral coherence terms generated by
$(\vec{x}_A\otimes\vec{x}_B)(\vec{x}_A\otimes\vec{x}_B)^\dagger$.

Let $t=\Delta\tau/2$ and $t'=\Delta\tau'/2$. The tensor product $\vec{x}_A\otimes\vec{x}_B$ generates 16 frequency-dependent phase components. Since the frequency degree of freedom is not resolved in the polarization measurement, these components should not be collected into a spectrally averaged state vector. Instead, the frequency trace gives a spectral coherence matrix $\Gamma\in \mathbb{C}^{16\times 16}$,
\begin{equation}
\Gamma
=
\int\int d\omega_A d\omega_B,
S(\omega_A,\omega_B)
\left[
\vec{x}_A(\omega_A)\otimes\vec{x}_B(\omega_B)
\right]
\left[
\vec{x}_A(\omega_A)\otimes\vec{x}_B(\omega_B)
\right]^\dagger .
\end{equation}
Each element of $\Gamma$ is obtained by applying $I(\Delta t_A,\Delta t_B)$ to the delay differences between the corresponding phase components.

Finally, defining
\begin{equation}
D=
m_A^{(H)}\otimes m_B^{(V)}
-
m_A^{(V)}\otimes m_B^{(H)},
\end{equation}
the reduced polarization density matrix after the CAPMF transformation is written as
\begin{equation}
\rho_o
=
\frac{1}{2}
D\Gamma D^\dagger .
\end{equation}
This expression is equivalent to Eq.~(\ref{eq9}), but separates the calculation into the polarization coefficient matrices $m_i^{(j)}$ and the spectral coherence matrix $\Gamma$. It therefore provides a convenient analytic form for evaluating the frequency-unresolved polarization state transmitted through the CAPMF.

\end{appendices}

\bmhead{Abbreviations}
QKD, quantum key distribution; PMF, polarization-maintaining fiber; SMF, single-mode fiber; HWP, half-wave plate; QWP, quarter-wave plate; PMD, polarization mode dispersion; PPKTP, periodically poled potassium titanyl phosphate; SPDC, spontaneous parametric down conversion; DPBS, dual-wavelength polarizing beam splitter; F, bandpass filter; R, phase retarder; DHWP, dual-wavelength half-wave plate; CW, clockwise; CCW, counter-clockwise; PBS, polarizing beam splitter; SPAD, single-photon avalanche diode; TCSPC, time-correlated single-photon counter; Pol, polarizer.

\bmhead{Acknowledgements}
The authors thank Heonoh Kim (KAIST) for his assistance in analyzing and resolving issues encountered during the experiments.

\bmhead{Author contributions}
J.-W.K. led the study and performed the theoretical modeling, experimental implementation, data analysis and manuscript preparation. M.K. proposed the initial concept, contributed to the experimental design and measurement procedure. J.P. contributed to the experimental implementation, data analysis. J.O., K.L., and B.-S.C. contributed to technical discussions. C.J.Y. provided project supervision and administrative support. All authors reviewed and approved the manuscript.

\bmhead{Funding}
This work was supported by the Institute of Information \& communications Technology Planning \& Evaluation(IITP)
grant funded by the Korea government(MSIT) (RS-2022-II221014, RS-2024-00398716, RS-2019-II190005, RS-2025-02218080)


\bibliography{sn-bibliography}

\end{document}